\documentclass{elsarticle}

\usepackage{lineno,hyperref}
\usepackage{amsmath}

\journal{arXiv}

\begin{document}

\begin{frontmatter}

\title{Application of a Modified Harmony Search Algorithm in the Optimal Arrangement of a Novel Three Dimensional Multiphase Flow Imaging Device}

\author[HEU]{Haihang Wang}

\author[HEU]{He Xu\corref{mycorrespondingauthor}}
\cortext[mycorrespondingauthor]{Corresponding author}
\ead{railway\_dragon@sohu.com}
\ead{Tel:(+86)13351117608}
\author[Finland]{Xiaozhi Gao}
\author[HEU]{Zitong Zhao}
\author[HEU]{Jinwei Huang}

\address[HEU]{College of Mechanical and Electrical Engineering, Harbin Engineering University,	Harbin, China}
\address[Finland]{Department of Electrical Engineering and Automation, School of Electrical Engineering, Aalto University, Aalto, Finland}

\begin{abstract}
Gas-liquid two-phase flow is a typical flow, and bubble characteristic measurement is of great importance to study the flow mechanism and guide the practical fluid mechanical engineering. In this paper, a novel three dimensional (3D) multiphase flow imaging device was designed to measure the transparent object that has an opaque object in the center of the observed area. Its mathematical model was built and the constraints were defined based on the geometrical relationship and design requirements. A modified harmony search (HS) algorithm was integrated and applied to optimize the arrangement of the single-camera-multi-mirror device. As a case study, the 3D multiphase flow imaging method was applied in the the 3D reconstruction of the cavitation bubble cluster inside a water hydraulic valve. The statistics of the Pareto data shows the good performance of the modified HS algorithm. And the cavitation experimental results shows that the method is valid, and the cavitation bubble cluster can be reconstructed with quite high precision.
\end{abstract}


\end{frontmatter}

\section{Introduction}
Multiphase flow is a typical flow, and bubble characteristic measurement is of great importance to discover the flow mechanism and guide the practical fluid mechanical engineering \cite{ZHENG20105264}. With the development of computer and optoelectronic techniques, visual inspection \cite{HONKANEN201025, WU20132928} based on photography has been widely used in multiphase flow measurement. However, the overlapping or opaque objects, which would decrease the reconstruction accuracy of modality and motion feature parameters of bubbles in gas-liquid two-phase, exist widely and cannot be recognized effectively. From the past studies, virtual binocular stereo vision has been applied by Xue et al. \cite{Xueting, Xueting2}, who did an experiment by using one camera and 4 mirrors to match and reconstruct bubble trajectory motion in a glass-made water tank from two direction. The mirror-based approach was applied in numerous single-camera stereovision techniques in various fields \cite{YU2016120, PAN201625, YU201717}. And Pan et al. optimized the parameters of the proposed 3D-DIC system, including the 3D position and orientation of the camera, effective focal length, principle point coordinates, and lens distortion coefficients \cite{Pan44412}. Inspired the emerging studies of the single-camera mirror-based stereo image correlation technique \cite{Pan2018}, a 3D imaging device combining single camera and multi-mirror was presented for capturing the comprehensive multiphase flow images information with an opaque object in the center of the observation area. And its design arrangement is a multi-objective problem (MOP), which is necessary to be optimized.

For solving the multi-objective problem, there are a variety of available algorithms \cite{app8071144, app8020175}, including genetic algorithm (GA), particle swarm optimization algorithm, artificial fish swarm  algorithm, and harmony search algorithm. HS algorithm was proposed and established first in 2001 by Geem et al. \cite{Zong600201}, which simulated the process of improvisation music, that is, the music player adjusts the performance to a wonderful harmony state through impromptu music adjustment. Harmony algorithm in the application of MOP, the search for the optimal solution is rather similar to the production of offspring in GA with mutation and crossover operations. The decision variables in the target problem are similar to the pitch of each instrument. The HS algorithm is a heuristic global search algorithm which also has the advantages of less parameters and high efficient calculation \cite{WANG20102826}.

As for the application of harmony algorithms in the MOP problems, many researchers have applied it in various areas, such as the location, size and power factor of Distributed Generation \cite{RN1552}, the multi-objective flexible job shop scheduling problem \cite{RN1542}, the scheduling problem of hydraulic systems pump \cite{RN1539, RN1550}, the urban traffic light scheduling problem \cite{RN1541} and so on.
However, empirical studies have shown that the original harmony search algorithm in dealing with multi-objective constrained optimization problems is being subject to certain restrictions \cite{PAN2010830, RN1538}. To overcome these shortcomings, utilization of a modified harmony search algorithm for multi-objective optimization to optimize the shielding effectiveness of wheel for secondary development is necessary.

Dai et al. \cite{RN1538} improved HS to solve the trouble for novice users about the parameters which need to be set by users according to experience and problem characteristics. Amaya et al. \cite{RN1536} presented a novel modification of the Harmony Search (HS) algorithm which is able to self-tune as the search progress. Yuan et al. \cite{RN1557} integrated HS in the research of the weakness about parallel chaos optimization algorithm, which aims to obtain optimum solution accurately. Meysam Gheisarnejad \cite{RN1544} developed cuckoo optimization algorithm into HS algorithm to design a secondary controller for two practical models of load frequency control problem. Gao et al. \cite{gao2009uni} propose two modified HS methods to deal with the uni-modal and multi-modal optimization problems. The first employed a novel HS memory management approach to handle the multi-modal problems and the second utilized the Pareto-dominance technique, which targets at the constrained problems. The MOP of the single-camera-multi-mirror device concerns multiple constraint conditions. In this paper, the two modified HS methods were applied and integrated into one improved HS algorithm, which absorbs the advantages in both two HS methods.

\section{Optimization Algorithm}
\begin{figure}[htbp]
  \centering
  \includegraphics[width=10cm]{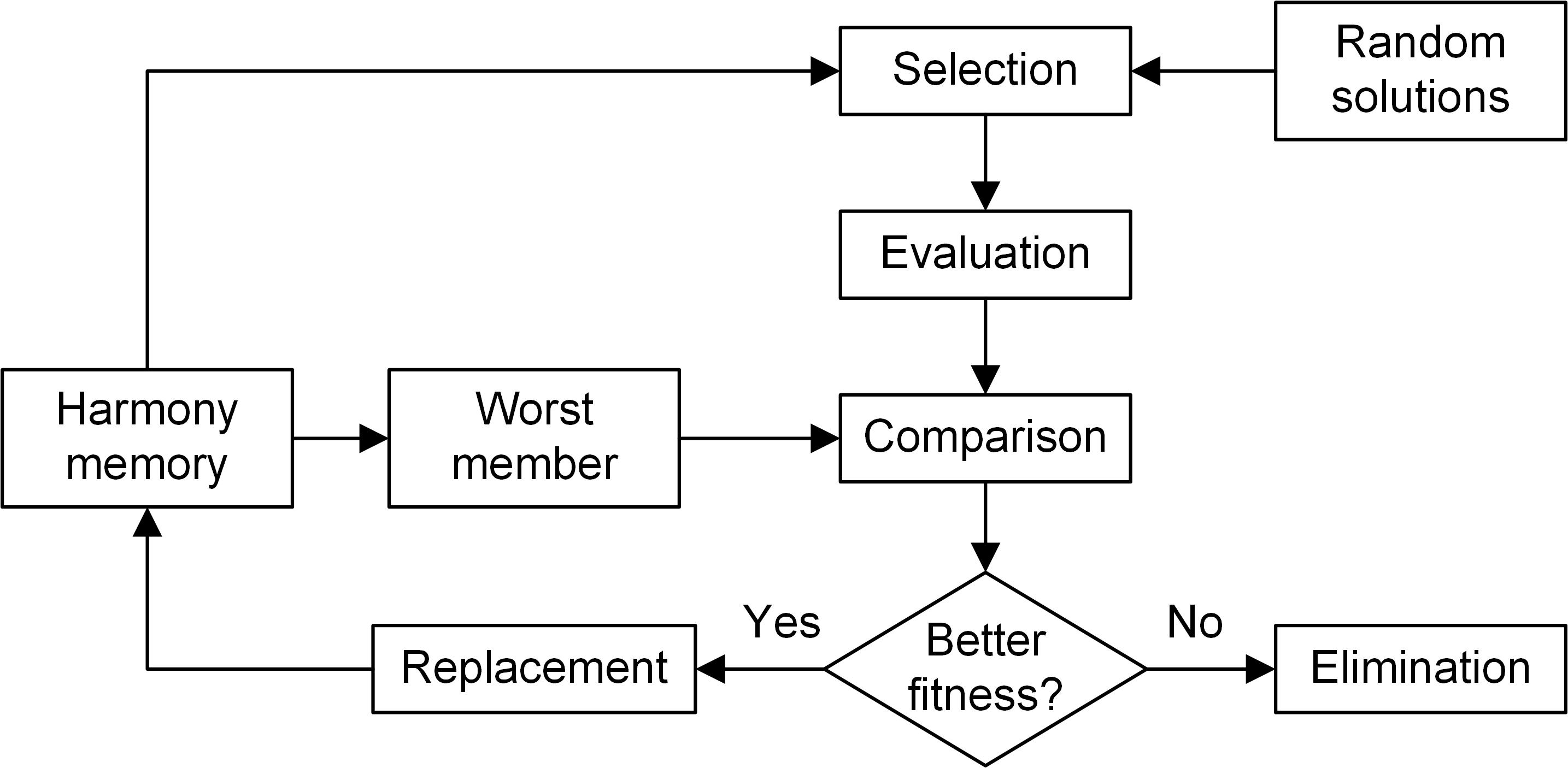}\\
  \caption{Flowchart of the basic HS method}\label{fig09}
\end{figure}

It is known that when experienced musicians compose a harmony, it is usually by trying various possible combinations of the music pitches stored in memory. This kind of effective search for a perfect harmony is analogous to the procedure of finding an optimal solution in engineering problems. The HS method is inspired by the working principles of the harmony improvisation \cite{Zong600201}. And as shown in Figure \ref{fig09} the flowchart of the basic HS method was presented.

The initial HS Memory (HM) consists of a given number of randomly generated solutions to the optimization problems under consideration. For a n-dimension problem, a HM with the size of HMS can be represented as follows:

\begin{equation}\label{eqHM}
  {\rm{HM = }}\left[ {\begin{array}{*{20}{l}}
{x_1^1,x_2^1, \cdots ,x_N^1}\\
{x_1^2,x_2^2, \cdots ,x_N^2}\\
 \vdots \\
{x_1^{HMS},x_2^{HMS}, \cdots ,x_N^{HMS}}
\end{array}} \right]
\end{equation}
where,$\left[ {x_1^i,x_2^i, \cdots ,x_n^i} \right]$$(i = 1,2, \cdots ,HMS)$is a candidate solution.

The basic harmony search method is difficult to find optimum result of multi-objective problem because in the iterative process, the harmony memory members easily stop in local optimal solution. Therefore, it is important to maintain the diversity of HM.

Assuming the current HM fitness value is expressed as ${f_i}(i = 1,2, \cdots ,HMS)$ and after a new candidate solution $\left( {{{x'}_1},{{x'}_2}, \cdots ,{{x'}_n}} \right)$ as fitness, then the distance ${d_i}\left( {i = 1,2, \cdots ,HMS} \right)$ between it and all other HMs is expressed as following equation:
\begin{equation}\label{eqdi}
    {d_i} = \left\| {\left[ {{{x'}_1},{{x'}_2}, \cdots ,{{x'}_n}} \right] - \left[ {x_1^i,x_2^i, \cdots ,x_n^i} \right]} \right\|
\end{equation}

Besides, the average fitness value of HM is expressed as follow:
\begin{equation}\label{eqF}
   \overline F  = \frac{{\sum\limits_{i = 1}^M {{f_i}} }}{M}
\end{equation}
where, $M$ is the number of the HM members.

Though for the concept of ${d_i}\left( {i = 1,2, \cdots ,HMS} \right)$  must prevent the excessive similarity between the members from HM. In other word, modified HS would be suitable for handling multi-objective problem by maintaining the diversity of HM. Most of the practical optimization problems are in fact constrained optimization problems, whose goal is to find the optimal solution that satisfies a set of given constraints \cite{Miettinen2008, CHEN20113336}. In general, a constrained optimization problem is described as follows:

Find $ \mathbf{x} = \left( {{{x'}_1},{{x'}_2}, \cdots ,{{x'}_n}} \right)$ to satisfy:
\begin{equation}\label{eqminf}
  \begin{array}{l}
minf(\mathbf{x})\\
{\rm{s}}{\rm{.t}}{\rm{.}}\left\{ \begin{array}{l}
{g_i}(\mathbf{x}) \le 0,i = 1,2,3 \cdots I\\
{h_j}(\mathbf{x}) = 0,j = 1,2,3 \cdots J
\end{array} \right.
\end{array}
\end{equation}
where, $f(\mathbf{x})$ is the objective function, $g_i(\mathbf{x})\leq0$ and $h_j(\mathbf{x})=0$ are the inequality and equality constraint functions respectively. As a matter of fact that the equality constraint functions can be easily transformed into the inequality constraint functions:
$$|h_j(\mathbf{x})|\leq\varepsilon$$
where, $\varepsilon$ is a small enough tolerance parameter. Therefore, we only consider the inequality constraint functions $g_i(\mathbf{x})\leq0$, $i = 1,2,3 \cdots,I$. Since the constraint functions could divide the whole search space into some disjoint islands such problems are generally difficult to deal. Numerous constraint-handling techniques have been investigated during the past decades. One popular solution is to define a new fitness function $F(\mathbf{x}))$ to be optimized. For example, $F(\mathbf{x}))$ is the combination of the objective function $f(\mathbf{x}))$ and weighted penalty terms $P(\mathbf{x}))$, $i = 1,2,3 \cdots I$, which reflect the violation of the constraint functions:
\begin{equation}\label{eqFx}
  F({\rm{x}}) = f(x) + \sum\limits_{i = 1}^I {{w_i}} {P_i}(x)
\end{equation}
where $w_i$, $i = 1,2,3 \cdots I$, are the preset weights. The overall optimization performance depends on the penalty terms and their weights, and may significantly deteriorate with inappropriately chosen ones.

Besides, the HM only stores the feasible solution candidates. The new HM members are generated either from the existing HM members or in a random way. Nevertheless, those are not guaranteed to always meet all the constraints. In the original HS method, the new HM members satisfying the constraints can be obtained based on only trial and error method which may lead to a time consuming procedure particularly in the case of multiple and complex constraint functions.

In this improved HS algorithm, the advantage of those HM members that do not even meet the constraints is taken. The key issue is how to rank the HM members according to their objectives as well as constraint functions values. The HM members are divided into two different parts: feasible members and infeasible members. The former satisfy all the constraint functions while the latter do not. The ranking of the feasible HM members is straightforward that means they can be sorted using their objective functions values. However, for the infeasible ones, the ranking is based on the Pareto dominance of these HM members \cite{COELLOCOELLO2002193, Mateo2012}. An infeasible HM member dominates another, if none of its constraint functions values are larger and at least one is smaller. Formally, the Pareto dominance is defined as follows.

Suppose there are two infeasible HM members $\mathbf{x}^a$ and $\mathbf{x}^b$. If
$\forall i \in \{1,2,\cdots, I\}$, $g_i(\mathbf{x}^a)\leq g_i(\mathbf{x}^{b})\bigcap \exists i \in \{1,2,\cdots, I\}$, $g_i(\mathbf{x}^a)< g_i(\mathbf{x}^{b})$, it can be concluded that $\mathbf{x}^a$ dominates $\mathbf{x}^b$. For each infeasible HM member, we can calculate the number of the others that dominate it. That implies its relative degree of violation of the constraint functions. It is the ranking of an infeasible HM member and is determined by the number of other infeasible HM members by which it is dominated.

Once after the whole HM has been ranked, the worst HM member $\mathbf{x}^\#$ can be selected and compared with the new solution candidate $\mathbf{x}^*$. Note, $\mathbf{x}^*$ does not need to be feasible. When $\mathbf{x}^\#$ is compared with $\mathbf{x}^*$, $\mathbf{x}^*$ will replace $\mathbf{x}^\#$ only in one of the following three cases:

\textbf{Case 1}: $\mathbf{x}^*$ is feasible, and $\mathbf{x}^\#$ is infeasible.

\textbf{Case 2}: Both $\mathbf{x}^*$ and $\mathbf{x}^\#$ are feasible and $f(\mathbf{x}^*)<f(\mathbf{x}^\#)$.

\textbf{Case 3}: Both $\mathbf{x}^*$ and $\mathbf{x}^\#$ are infeasible and $\mathbf{x}^*$ dominates $\mathbf{x}^\#$.

\begin{figure}[htbp]
  \centering
  \includegraphics[width=12cm]{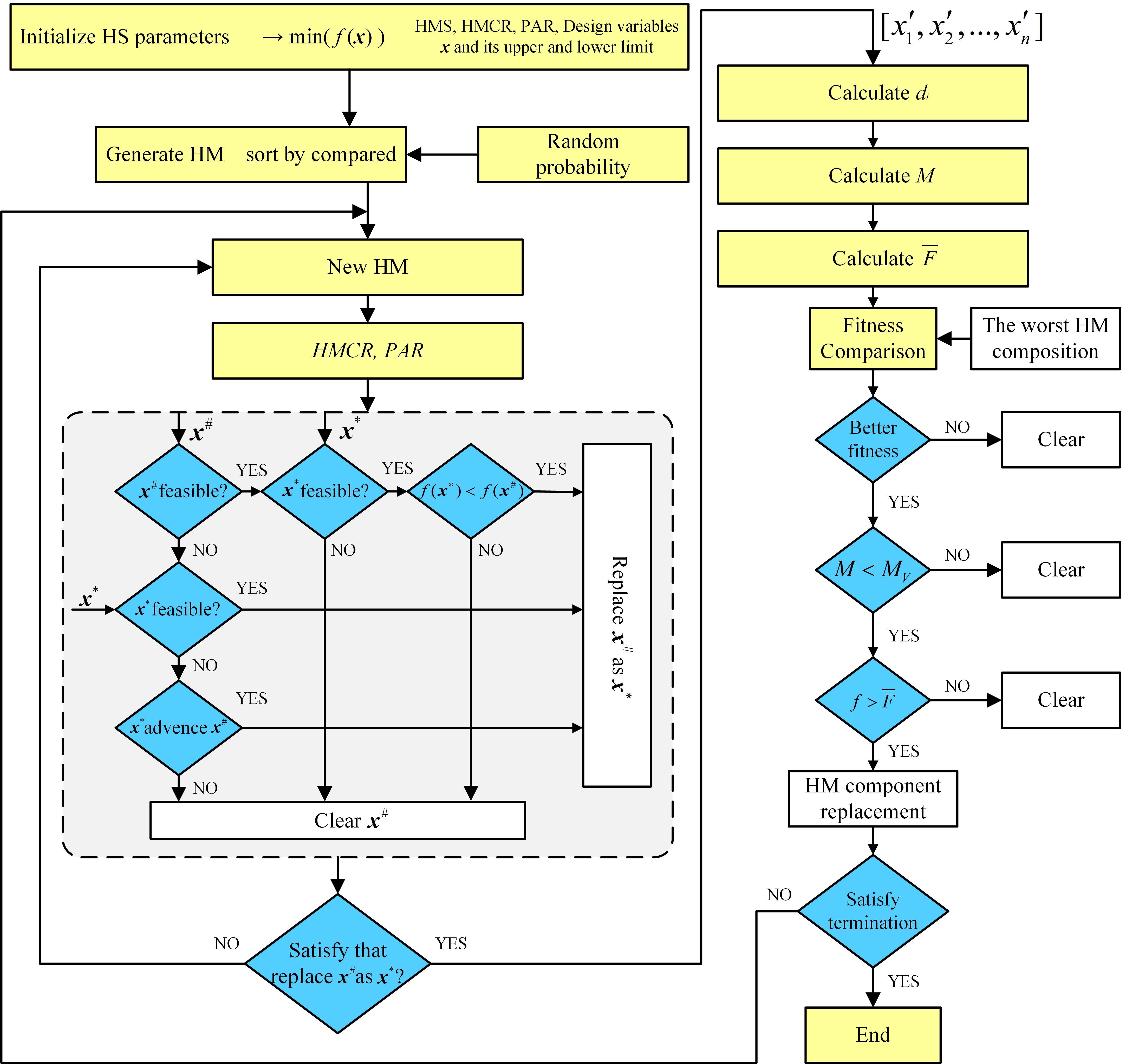}\\
  \caption{ Multi-objective optimization flowchart based on a modified HS algorithm}\label{fig03}
\end{figure}

The process is shown in Figure \ref{fig03}.
\section{Design Expression of the Single-camera-multi-mirror Device}
\subsection{Measurement Principle of the 3-D Imaging}

\begin{figure}[htbp]
  \centering
  \includegraphics[width=8cm]{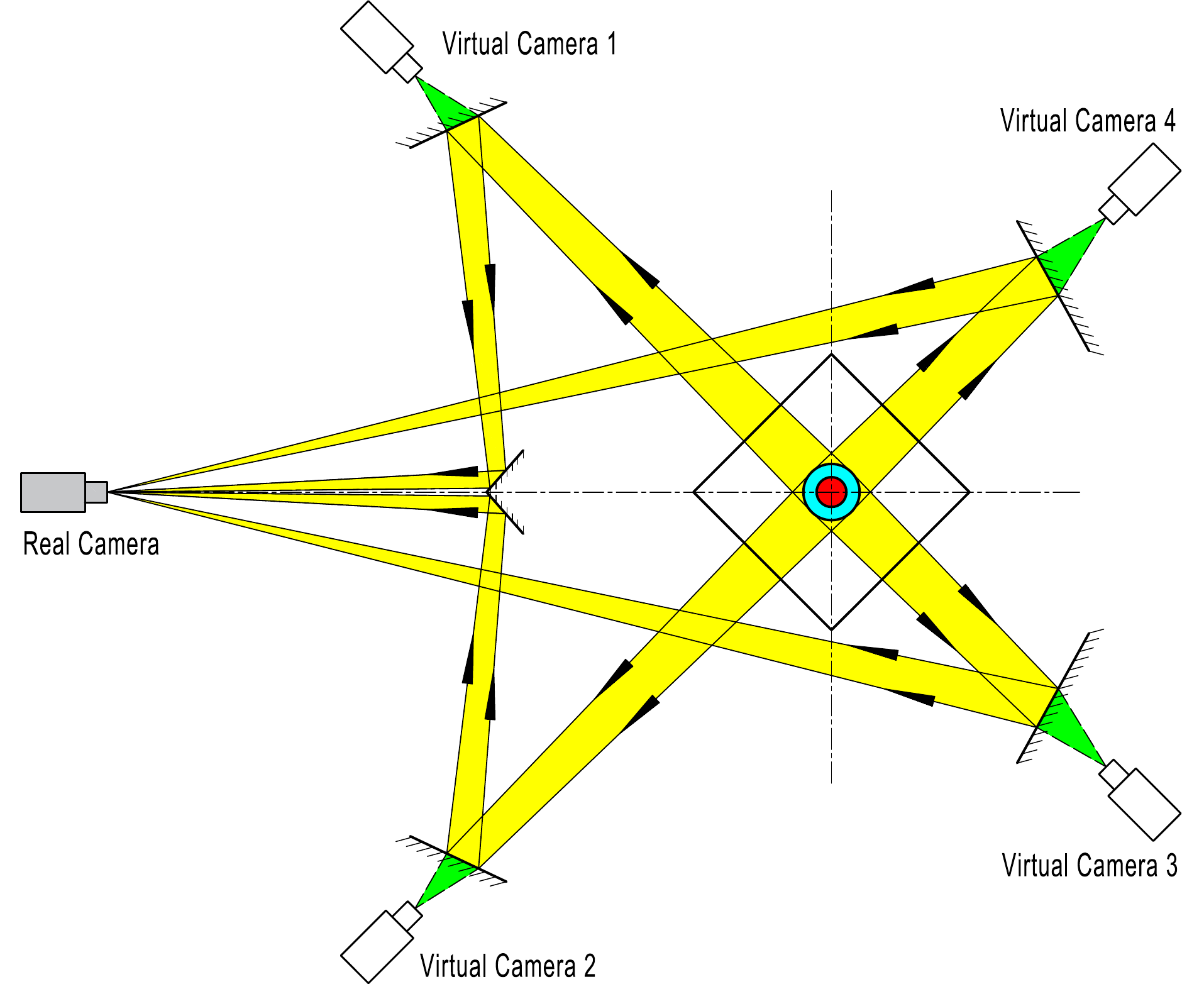}\\
  \caption{Schematic diagram of the 3-D imaging principle}\label{fig02}
\end{figure}

Figure \ref{fig02} shows the arrangement of the single-camera-multi-mirror device from the top view. The square the outer contour of the object being observed. The blue circle is the area that needs to be photographed. And the red circle is the area that was opaque. Using a single camera from one direction cannot capture the comprehensive image information of the blue area, because of the opaque object. To solve this problem, a novel single-camera-multi-mirror device was designed to enable one camera to capture the images from four vertical faces of a cuboid. The images of the 4 faces were reflected by the plane mirrors and focused on the camera lens.

To eliminate the image refringence caused by the oblique angle between the observed face and the virtual camera, the light center axes of the three virtual cameras were all defined to be perpendicular to the observed faces. Besides, the distance of the optical path of the four virtual cameras were set to be equal to ensure the consistent image resolution from the four sides and enable the lens to focus to the four observing areas.

\subsection{Mathematical Model of the Single-camera-multi-mirror Design}
\begin{figure}[htbp]
  \centering
  \includegraphics[width=12cm]{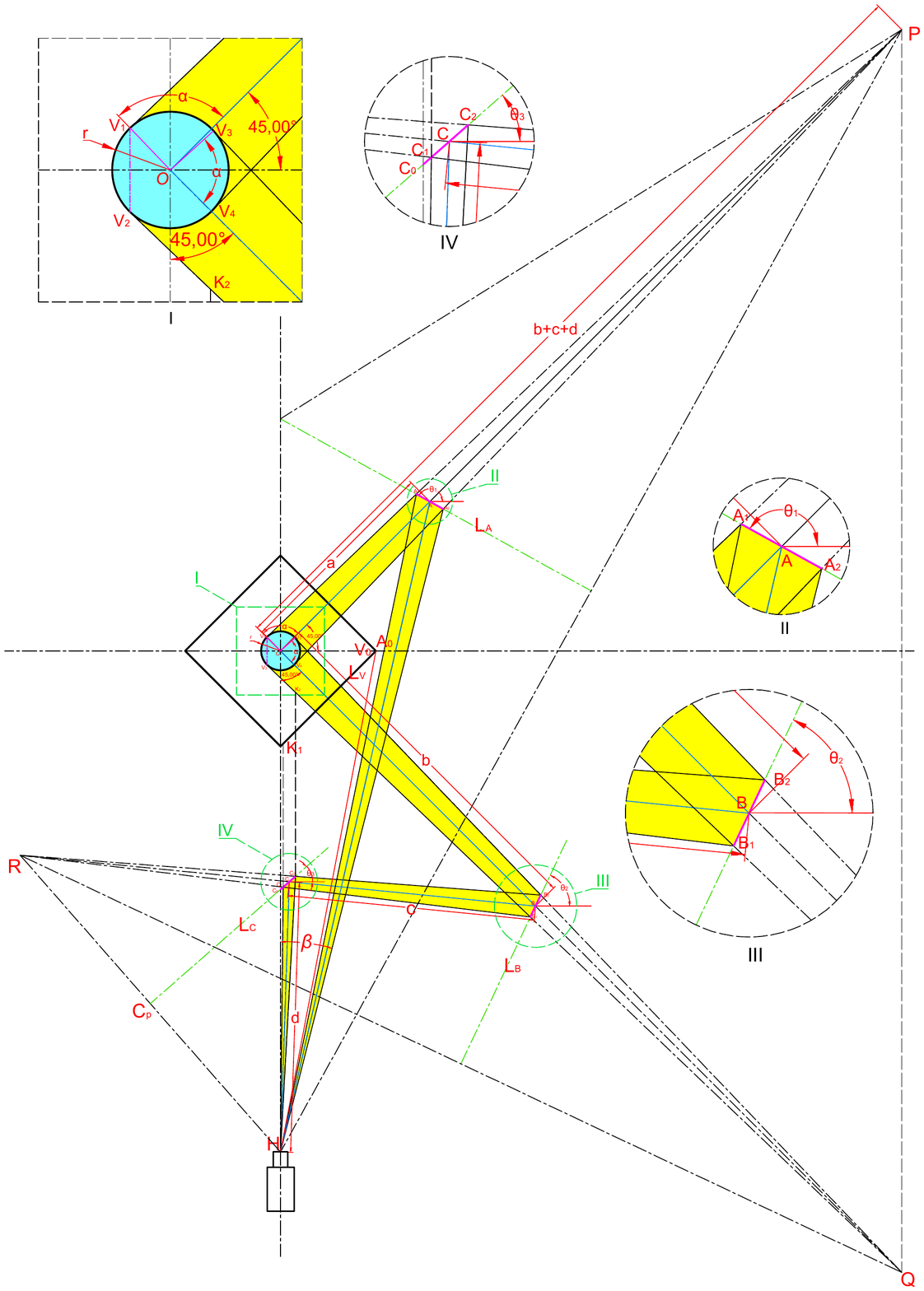}\\
  \caption{Establishment and parameter setting of the mathematical model}\label{fig01}
\end{figure}

To optimize the arrangement of the high-speed camera and six plane mirrors to build the mathematical model for optimization, the design diagram was drawn, as shown in Figure \ref{fig01}. The point $P$ and $Q$ are the position of the virtual camera 3 and 2  in Figure \ref{fig02}.
Due to the symmetrical relationship of the virtual camera 1  and 2, the optimal design of the two mirrors of virtual camera 1 and 4 was omitted. So the optimization design variables in practice were the position parameters of the camera (point $H$) and the three mirrors (marked as A, B and C). The radius of the observing area is defined as $r$. The length of the outline of the observed cuboid is defined as $l_V$. So, the coordinate of the point $V_0$ is $(\frac{l_V}{\sqrt{2}},0)$. The default unit in this paper is mm.

The coordinates of the virtual camera 2 and 3 are expressed as follows:
%
$ Q(\frac{b+c+d}{\sqrt{2}},-\frac{b+c+d}{\sqrt{2}}) $,
%
$ P(\frac{b+c+d}{\sqrt{2}},\frac{b+c+d}{\sqrt{2}}) $.
Besides, the coordinates of the plane mirrors A and B can be expressed as
$ A(\frac{b}{\sqrt{2}},\frac{b}{\sqrt{2}}) $, $ B(\frac{b}{\sqrt{2}},-\frac{b}{\sqrt{2}}) $.

The three angle optimization variables are defined as follows $\theta_1$ (mirror A), $\theta_2$ (mirror B) and $\theta_3$ (mirror C). The equations for the lines crossing the A and B mirrors are respectively dissected as follows:
$$ L_A: y=x\tan\theta_1+\frac{a}{\sqrt{2}}(1-\tan\theta_1) $$
$$ L_B: y=x\tan\theta_2-\frac{b}{\sqrt{2}}(1+\tan\theta_1) $$

As point $P$ and point $H$ are symmetrical about line $L_A$, the coordinates of $H$ are obtained as follows:
\begin{align*}
x_H&=x_P+y_P\tan\theta_1-y_H\tan\theta_1\\
y_H&=\frac{ 2x_P\tan\theta_1+y_P(\tan^2\theta_1-1)+\sqrt{2}a(1-\tan\theta_1)  }
    {1+\tan^2\theta_1}
\end{align*}

Due to $x_P=y_P$ and $x_H=0$, the following equation is obtained:
%
$$ d=-b-c+a\frac{ 2(\tan\theta_1-\tan^2\theta_1)}
    {1+2\tan\theta_1 -\tan^2\theta_1} $$

Similarly, because point $Q$ and point $R$ are symmetrical about line $L_B$, the coordinates of $R$ are obtained as follows:
%
\begin{align*}
x_R&=x_Q+y_Q\tan\theta_2-y_R\tan\theta_2 \\
y_R&=\frac{2x_Q\tan\theta_2+y_Q(\tan^2\theta_2-1)-\sqrt{2}b(1+\tan\theta_2)}
    {1+\tan^2\theta_2}
\end{align*}

The equation of line $L_{BR}$ can be obtained from two points $B$ and $R$.
%
$$ L_{BR}: y=\frac{y_B-y_R}{x_B-x_R}(x-x_R)+y_R $$

As point $C_P$ is the midpoint of the line segment $RH$, the coordinates of the point $C$ are $(\frac{x_R+x_H}{2},\frac{y_R+y_H}{2})$ and the equation of line $L_C$ can be expressed as:
$$ L_C: y=\tan\theta_3(x-\frac{x_R+x_H}{2})+\frac{y_R+y_H}{2} $$

As the point $C$ is the intersection of Line $L_C$ and Line $L_{BR}$, the coordinates of the point $C$ are obtained as follows:
\begin{align*}
x_C&=\frac{ \frac{y_B-y_R}{x_B-x_R}x_R-y_R-\frac{x_R+x_H}{2}\tan\theta_3+\frac{y_R+y_H}{2} }
    { \frac{y_B-y_R}{x_B-x_R}-\tan\theta_3 }\\
y_C&=\frac{y_B-y_R}{x_B-x_R}(x_C-x_R)+y_R
\end{align*}

In addition, as the line $L_C$ is perpendicular to $L_{BR}$, the following equation can be obtained:

\begin{equation}\label{eq154}
  \tan\theta_3\cdot\frac{y_R-y_H}{x_R-x_H}=-1
\end{equation}

Based on the coordinates of the point $B$ and $C$, the geometrical relationship must be met:

\begin{equation}\label{eq159}
  c=|BC|=\sqrt{(x_B-x_C)^2+(y_B-y_C)^2}
\end{equation}

The relative coordinate values of the points of $B$, $C$, $R$ and $H$ can be all derived, when the design parameters $a$, $b$, $c$, $\theta_1$, $\theta_2$ and $\theta_3$ are given. So, the $\theta_2$ and $\theta_3$ can be obtained by solving the function set of Equation \ref{eq154} and \ref{eq159}.
\begin{equation}
\left\{
\begin{split}
&x_H-x_R+\tan\theta_3(y_R-y_H)=0\\
&x_C(x_C-2x_B)+y_C(y_C-2y_B)+b^2-c^2=0
\end{split}
\right.
\end{equation}

In above, the optimization variables are $a$, $b$, $c$ and $\theta_1$. All the position and angles can be derived by giving a set of values of these 4 variables.

\subsection{Optimization Model}
The optimal arrangement of the 3D imaging device in this paper is a multi-objective constraint optimization problem. The resolution of the experiment videos was limited by the capability of the high-speed camera and the distance from the observed field to the camera lens. The resolution capability of the camera and the lens used in the experiment are fixed. The cavitation bubbles in the valve were quite small. In order to ensure the resolution of the bubble images, the distance of the optical path were optimized as short as possible. Then, the first optimization objective can be expressed as follows:
\begin{equation}\label{eqf1}
  f_1(\mathbf{x})=b+c+d
\end{equation}

Besides, the overall size of the 3D imaging device should be as compact as possible. The vertical length is defined as the second optimization objective:
\begin{equation}\label{eqf2}
  f_2(\mathbf{x})=x_B
\end{equation}
And the lateral length is defined as the third optimization objective:
\begin{equation}\label{eqf2}
  f_3(\mathbf{x})=y_A-y_H
\end{equation}

The objective function is mathematically defined by:
\begin{equation}
\min J(\textbf{x})=\min(f_1(\mathbf{x}),f_2(\mathbf{x}),f_3(\mathbf{x}))
\end{equation}
Subject to:
$$G(\mathbf{x})\leq0 $$
where $\textbf{x}=[a,b,c,\theta_1]^T$.

\subsection{Model Constraint}
The calculation of the following parameters was to define the constraint conditions to meet the geometrical requirements of the single-camera-multi-mirror design and ensure no interference between the optical paths.

To prevent the optical image reflected by the mirror A from being interrupted by the valve, there should be a certain interval between $A_0$ and $V_0$, as expressed in Equation \ref{eqxA0}:
\begin{equation}\label{eqxA0}
  x_{A_0}-x_{V_0}>3
\end{equation}

The left boundary of the optical path reflected by the mirror C is on the positive side of the $y$ axis, in case of influencing the mirror belonging to the virtual camera 1, which is symmetrical with the mirror C. So, the $x$ coordinate of the point $C_1$ should meet the following constraint:
\begin{equation}
  x_{C_1}>2
\end{equation}

The slope of the line $HC_2$ and $HV_0$ can be expressed as:
$$ k_{HC_2}=\frac{y_H-y_{C_2}}{x_H-x_{C_2}},\ k_{HV_0}=\frac{y_H-y_{V_0}}{x_H-x_{V_0}} $$
To prevent the optical path from the mirror A and C from interfering with each other, the slopes angle of the line $HC_2$ and $HA_1$ should meet:
\begin{equation}
\arctan k_{HC_2}-\arctan k_{HV_0}>1^\circ
\end{equation}

To prevent interference between the placement of the mirror C and the valve body, the position of the points $C_1$ and $C_2$ should meet:
\begin{equation}
    y_{C_1}-y_{K_1}<-30
\end{equation}
\begin{equation}
    y_{C_2}-y_{K_2}<-10
\end{equation}

The visual scope is limited by the viewing angle of the camera lens. So,
\begin{equation}
   \beta<17.5^\circ
\end{equation}
Above all, the constraint can be defined as:
$$G(\mathbf{x})=\{g_1(\mathbf{x}),g_2(\mathbf{x}),g_3(\mathbf{x}),g_4(\mathbf{x}),g_5(\mathbf{x}),g_6(\mathbf{x})\}^T$$
\noindent where,
\begin{equation*}
\left\{
\begin{split}
&g_1(\mathbf{x})=\arctan k_{HV_0}-\arctan k_{HC_2}+1^\circ\\
&g_2(\mathbf{x})=x_{V_0}-x_{A_0}+3\\
&g_3(\mathbf{x})=2-x_{C_1}\\
&g_4(\mathbf{x})=y_{C_1}-y_{K_1}+30\\
&g_5(\mathbf{x})=y_{C_2}-y_{K_2}+10\\
&g_6(\mathbf{x})=\beta-17.5^\circ\\
\end{split}
\right.
\end{equation*}
\section{Case Study}
\subsection{3D multiphase flow imaging of a water hydraulic valve}
In this section, the optimization arrangement of the 3D imaging device was applied in the multiphase flow reconstruction of the 3D cavitation bubble cluster in a water hydraulic valve. The main materials of the valve body is a type of transparent thermoplastic called Polymethylmethacrylate (PMMA, Perspex, acrylic glass). The two basic structural dimensions is given as $r=17$ and $l_V=118$ (unit: mm). Besides, there is the valve core in the center of the observed area, which obscured quite much sight. Because of the inevitable space for connecting the inlet/oulet hydraulic pipes, there is no space for the mirror in practice, and the virtual camera 4 was omitted.

The linear constraints of design parameters are defined according to the design requirements as follows.
\begin{equation}
\left\{
\begin{split}
&150<a<400\\
&150<b<400\\
&150<c<400\\
&145^\circ < \theta_1<180^\circ\\
\end{split}
\right.
\end{equation}

\begin{figure}[htbp]
  \centering
  \includegraphics[width=10cm]{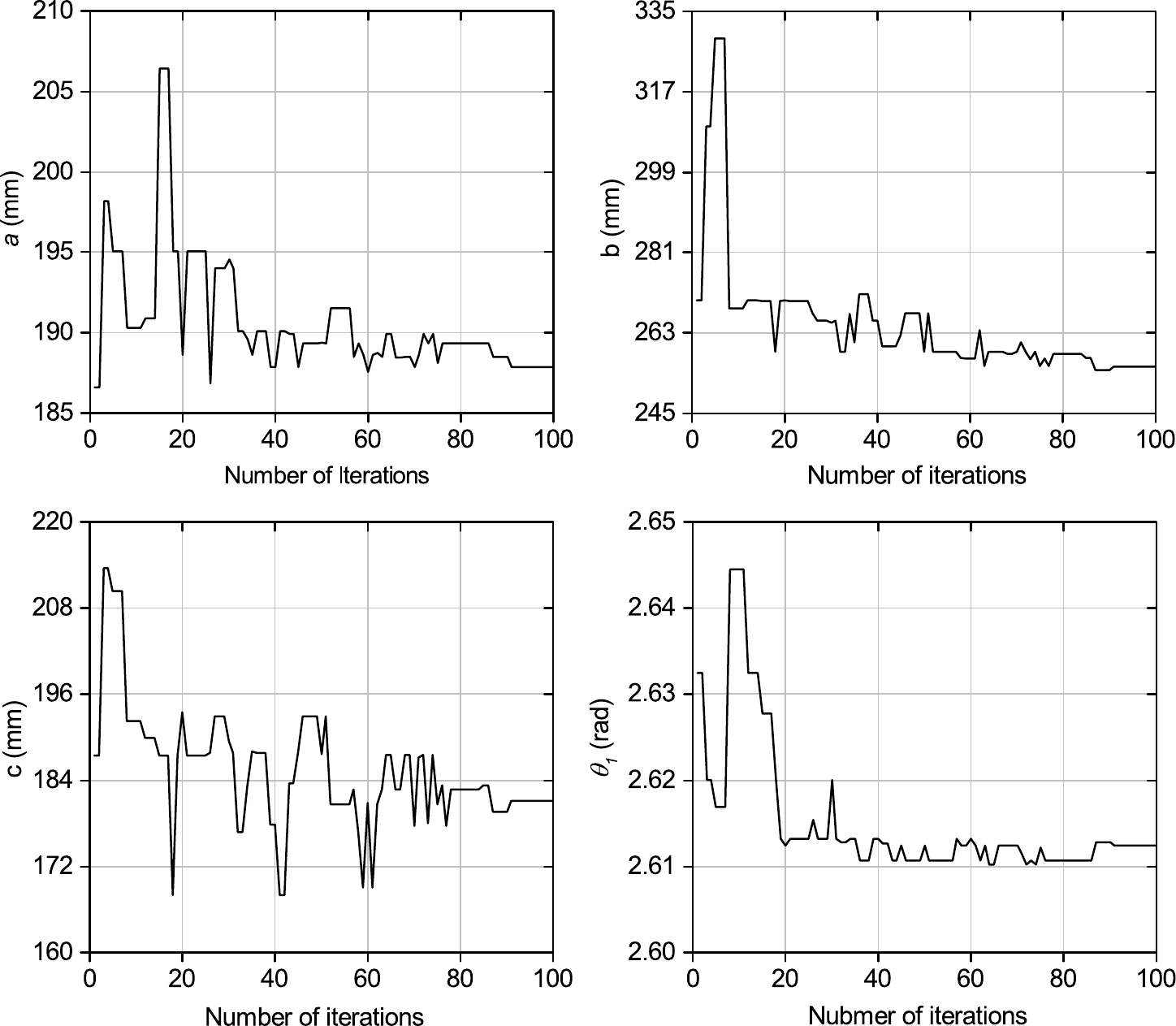}\\
  \caption{Optimization variables versus iterations during the optimization process}\label{fig10}
\end{figure}
\begin{figure}[htbp]
  \centering
  \includegraphics[width=10cm]{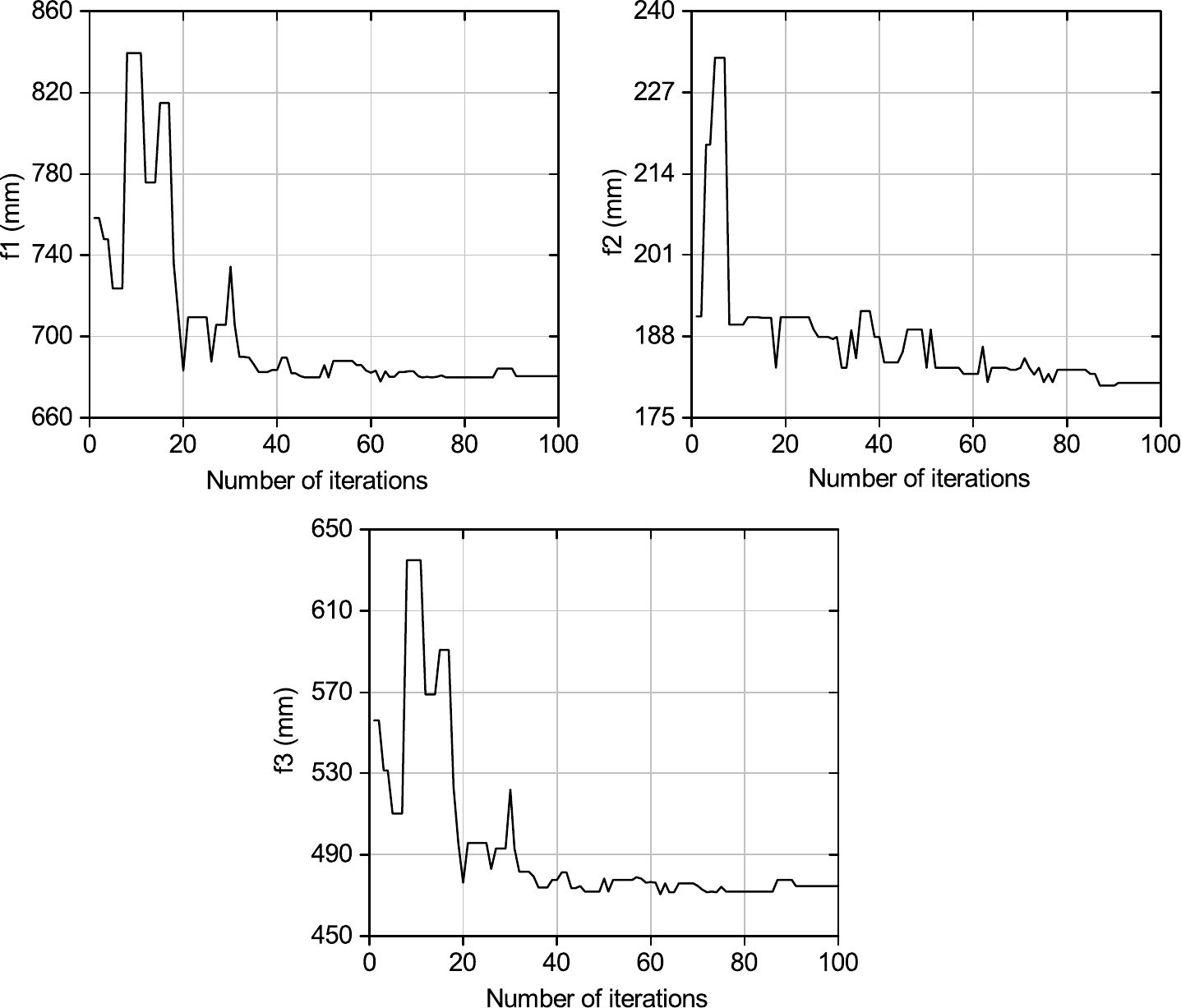}\\
  \caption{Objective functions versus iterations during the optimization process}\label{fig11}
\end{figure}

The relevant parameters in the HS algorithm are as follows: HMS=50, HMCR=0.75, PAR=0.4. The number of the Pareto solution is equal to 10. The number of iterations is 100 times, while 20 new sets of the optimization variables' solutions will replace the worse solutions in each iteration.
The changes of the variables and objective functions during the optimization procedure are illustrated in Figures \ref{fig10} and \ref{fig11}. As shown in Figure \ref{fig11}, in the first 20 iterations, the value of $f_1$, $f_2$ and $f_3$ were greatly fluctuate. After that, all the three objective functions steadily decreases and converges. This shows the efficiency of the algorithm in convergence speed. As a result, Table \ref{tab01} gives the Pareto results acquired.

\begin{table}[htbp]
  \small
  \centering
  \caption{Pareto resolutions of optimization}\label{tab01}
    \begin{tabular}{cccccccc}
    \hline
    No. & $a$/mm & $b$/mm  & $c$/mm  & $\theta_1$/rad & $f_1(\mathbf{x})$/mm  & $f_2(\mathbf{x})$/mm  & $f_3(\mathbf{x})$/mm  \\
    \hline
    1     & 187.879  & 255.392  & 181.091  & 2.612  & 680.596  & 180.589  & 474.469  \\
    2     & 188.501  & 254.701  & 179.621  & 2.613  & 684.255  & 180.101  & 477.499  \\
    3     & 189.329  & 257.592  & 184.673  & 2.610  & 678.156  & 182.145  & 470.137  \\
    4     & 189.329  & 255.584  & 177.641  & 2.611  & 679.927  & 180.725  & 471.979  \\
    5     & 189.329  & 257.440  & 183.339  & 2.611  & 679.927  & 182.038  & 471.979  \\
    6     & 189.329  & 257.338  & 183.339  & 2.611  & 679.927  & 181.966  & 471.979  \\
    7     & 189.185  & 260.852  & 181.046  & 2.610  & 677.640  & 184.451  & 469.780  \\
    8     & 189.329  & 257.195  & 178.050  & 2.611  & 679.927  & 181.864  & 471.979  \\
    9     & 189.185  & 260.852  & 181.046  & 2.610  & 677.640  & 184.451  & 469.780  \\
    10    & 189.185  & 260.852  & 181.046  & 2.610  & 677.640  & 184.451  & 469.780  \\
    \hline
    \end{tabular}%
  \label{tab:addlabel}%
\end{table}%

\begin{figure}
  \centering
  \includegraphics[width=12cm]{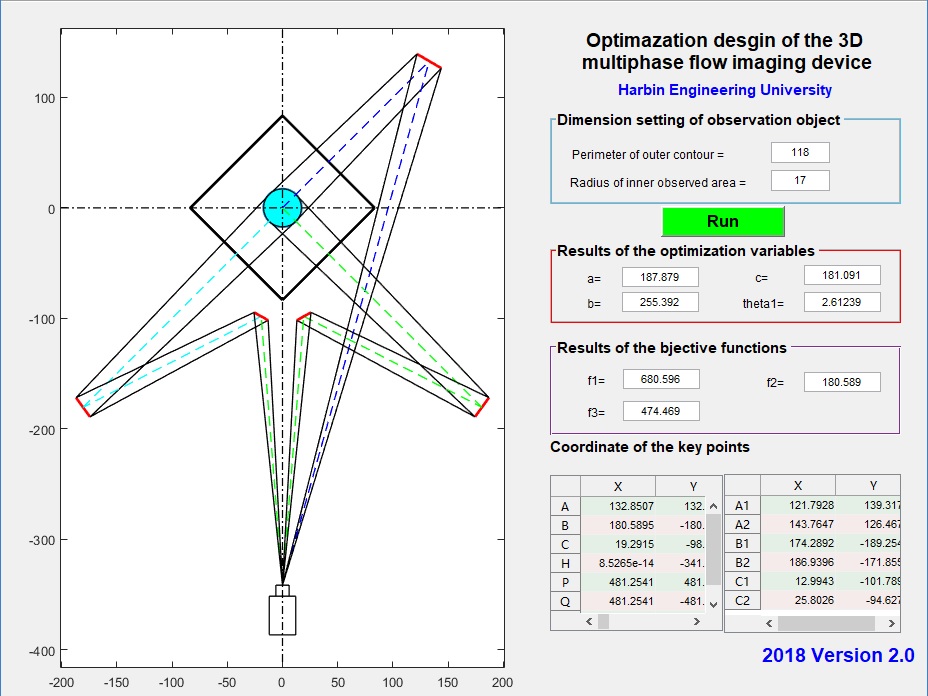}\\
  \caption{GUI of the optimizations design}\label{fig12}
\end{figure}

In order to facilitate other designers to apply the optimization method presented in this paper, a graphical user interface (GUI) was developed as shown in Figure \ref{fig12}, which integrated the functions of inputting the initial object parameters, running the modified HS algorithm, displaying the optimization results and drawing the optimal arrangement design. As a result, the optimal solution is obtained and selected as follows:
\begin{equation}
\begin{split}
\textbf{x}_{opt}=&[a_{opt},b_{opt},c_{opt},\theta_{1opt}]^T\\
=&[187.879, 255.392, 181.091, 149.679^\circ]^T
\end{split}
\end{equation}

\begin{figure}
  \centering
  \includegraphics[width=12cm]{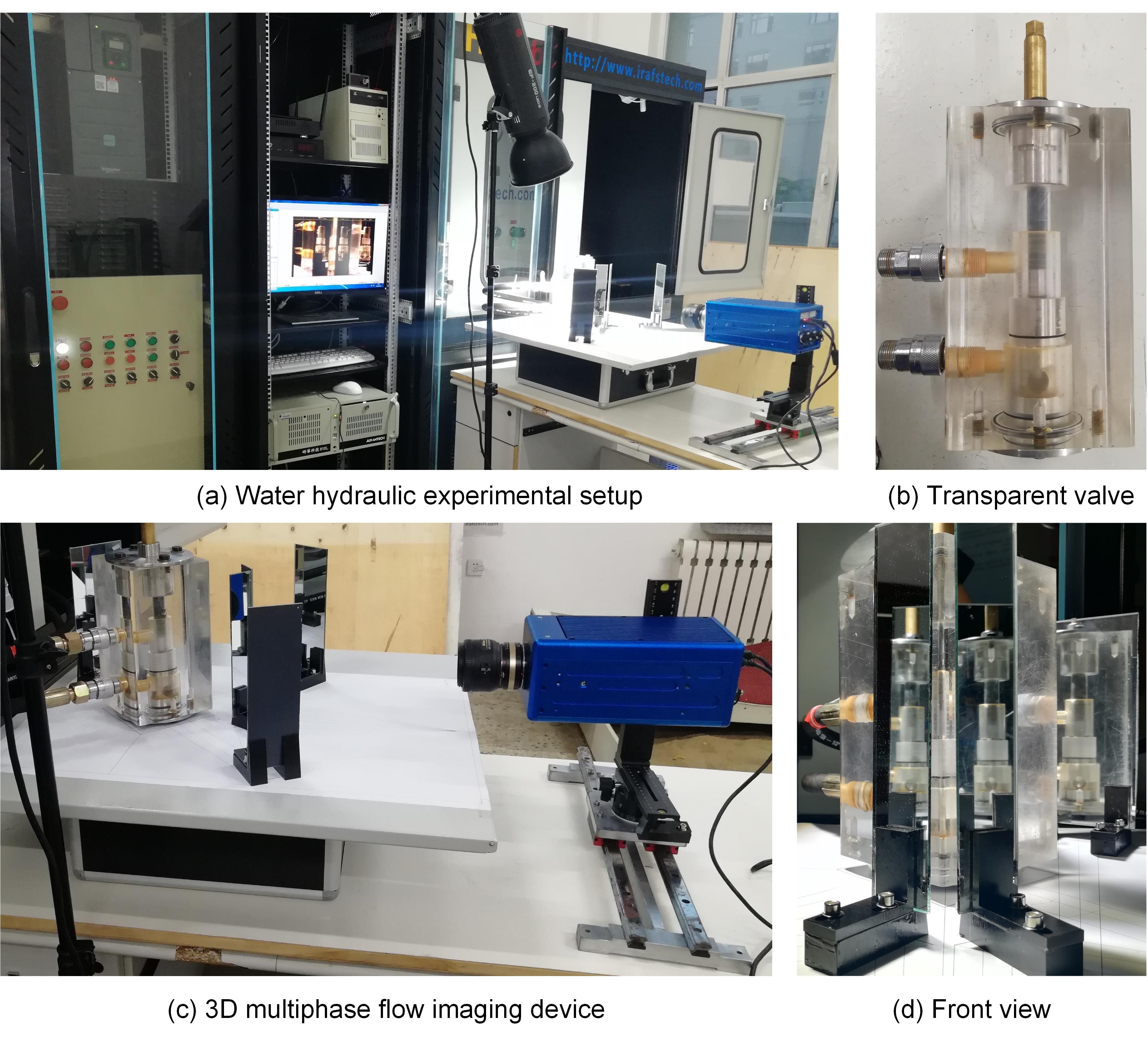}\\
  \caption{3D imaging experiment of the transparent water hydraulic valve}\label{fig04}
\end{figure}

Figure \ref{fig04} shows the photographs of a 3D imaging experiment. As shown in Figure \ref{fig04}(a), the water hydraulic experimental setup consisted of the water hydraulic platform, industrial personal computer (IPC), and the 3D imaging device. The transparent water hydraulic valve used as the observation object was shown in Figure \ref{fig04}(b). The arrangement design drawing was printed on an A2 sheet of paper. As shown in Figure \ref{fig04}(c). five plane mirror and a high-speed camera were placed based on the printed design paper. From the front view, the images from three sides of the valve was captured.
\subsection{Algorithm Performance}
To testify the good performance on locating most of the local optima in addition to the global optimum, the Pareto data of every iteration was collected. 2000 arrangement solutions of the 3D imaging device were recorded. The position of the mirror C were decided by the coordinate of the point $C$. Point $A$, $B$ and $H$ can only move on a certain line, while point C can move within a certain area.  And both $x_C$ and $y_C$ were decided by the optimal variables $a$, $b$, $c$,$\theta_1$, which can comprehensively reflect the performance. So we select point $C$ to analyze its position distribution and density.

\begin{figure}
  \centering
  \includegraphics[width=8cm]{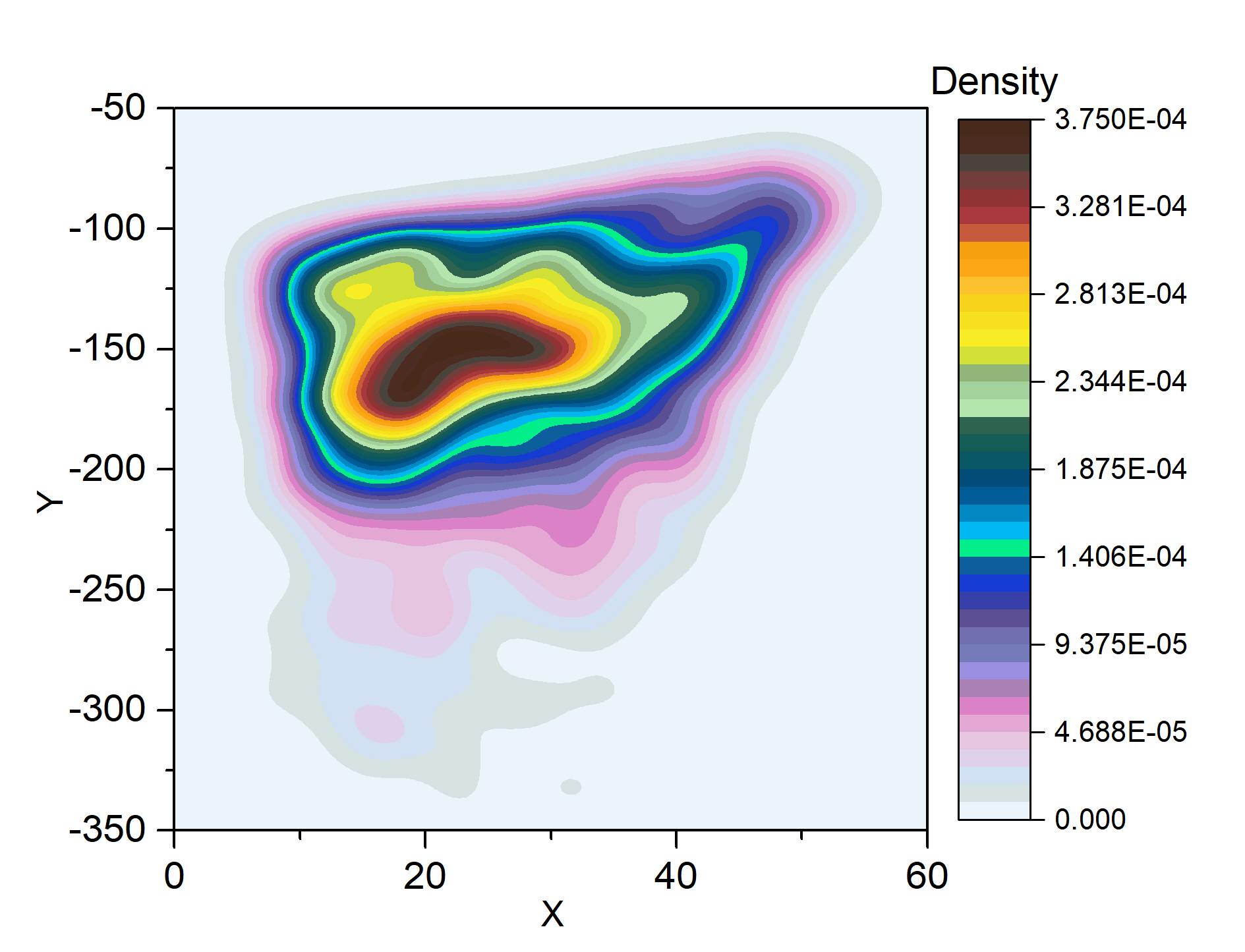}\\
  \caption{Kernel density of the point $C$}\label{fig08}
\end{figure}
Figure \ref{fig08} shows the density analysis of the point $C$. The 10 Pareto solutions were more concentrated in the pea-shaped area around the optimal solution.

\begin{figure}
  \centering
  \includegraphics[width=12cm]{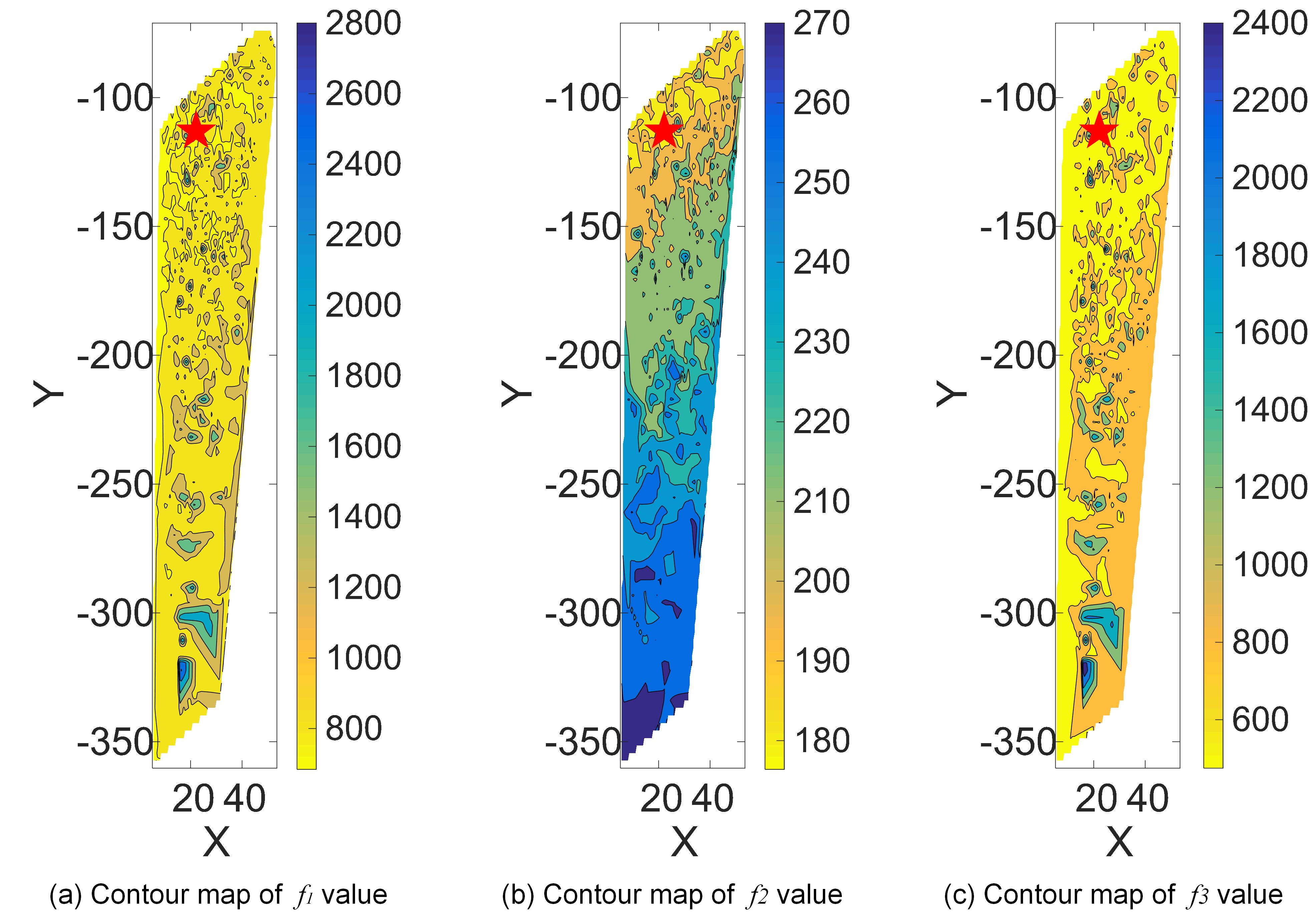}\\
  \caption{Position distribution of point $C$ and its optimization objectives' value}\label{fig07}
\end{figure}
Figure \ref{fig07} illustrates the contour maps of the optimization objectives' values about point $C$. The red star symbol is marked as the optimal position of the point $C$. The distribution of the point $C$ is quite extensive, which indicates that there were spread selections considered as the optimal solutions.
Compared Figure \ref{fig07}(b) with Figure \ref{fig07}(a) and (C), the area of the relative small value of the $f_2(\mathbf{x})$ is quite limited, which plays a more obvious role than $f_1(\mathbf{x})$ and $f_3(\mathbf{x})$. And the point $C$ optimal solution was at the lowest area of the $f_1(\mathbf{x})$ , $f_2(\mathbf{x})$ and $f_3(\mathbf{x})$.

\subsection{3-D Cavitation Bubble Cluster Reconstruction}
\begin{figure}
  \centering
  \includegraphics[width=9cm]{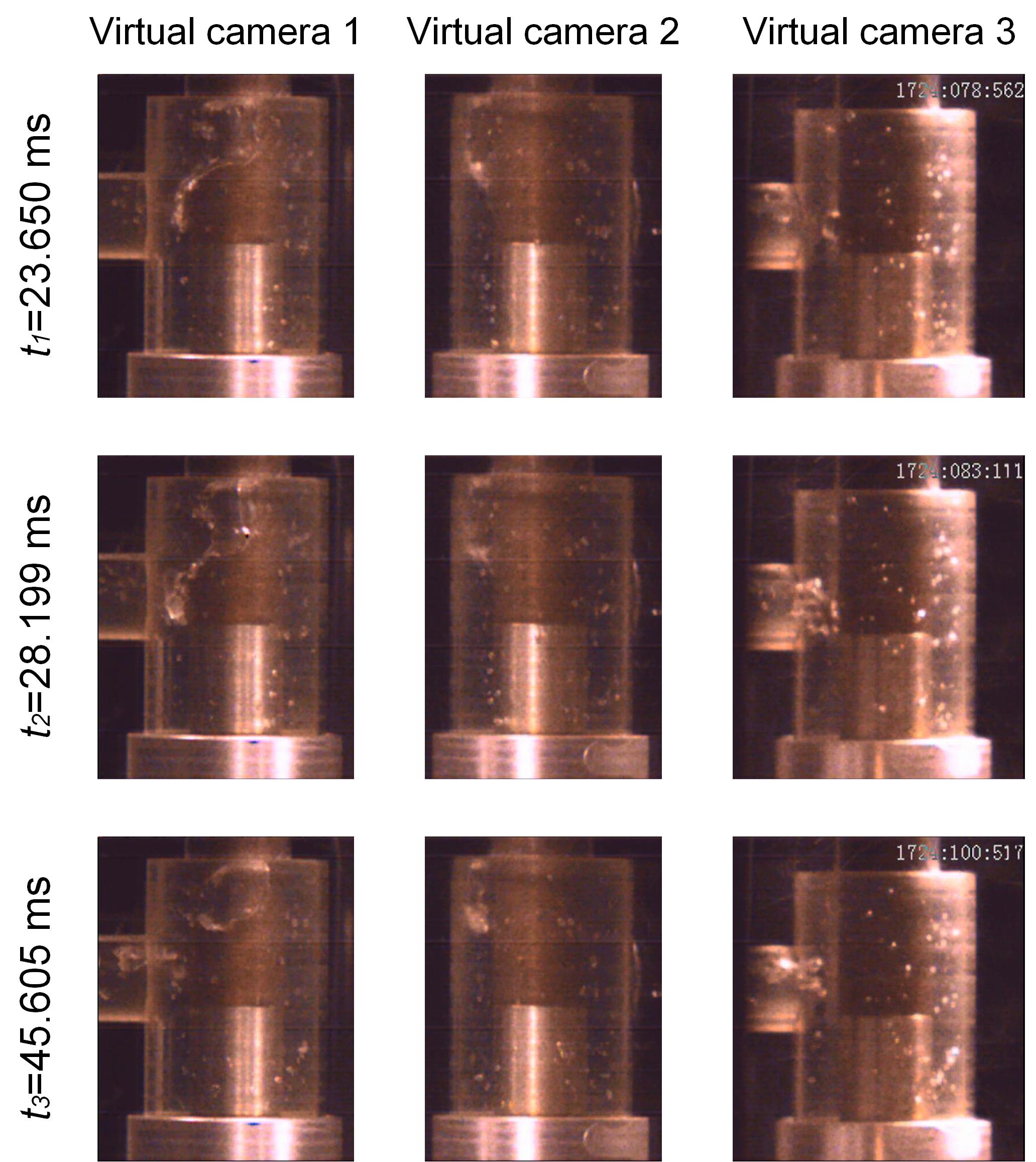}\\
  \caption{Experimental images in a short time}\label{fig05}
\end{figure}

Figure \ref{fig05} shows three randomly selected frames of the experimental video, whose time interval was only about 23 milliseconds (ms). The cavitation bubbles inside the valve were captured from the three sides.

\begin{figure}
  \centering
  \includegraphics[width=12cm]{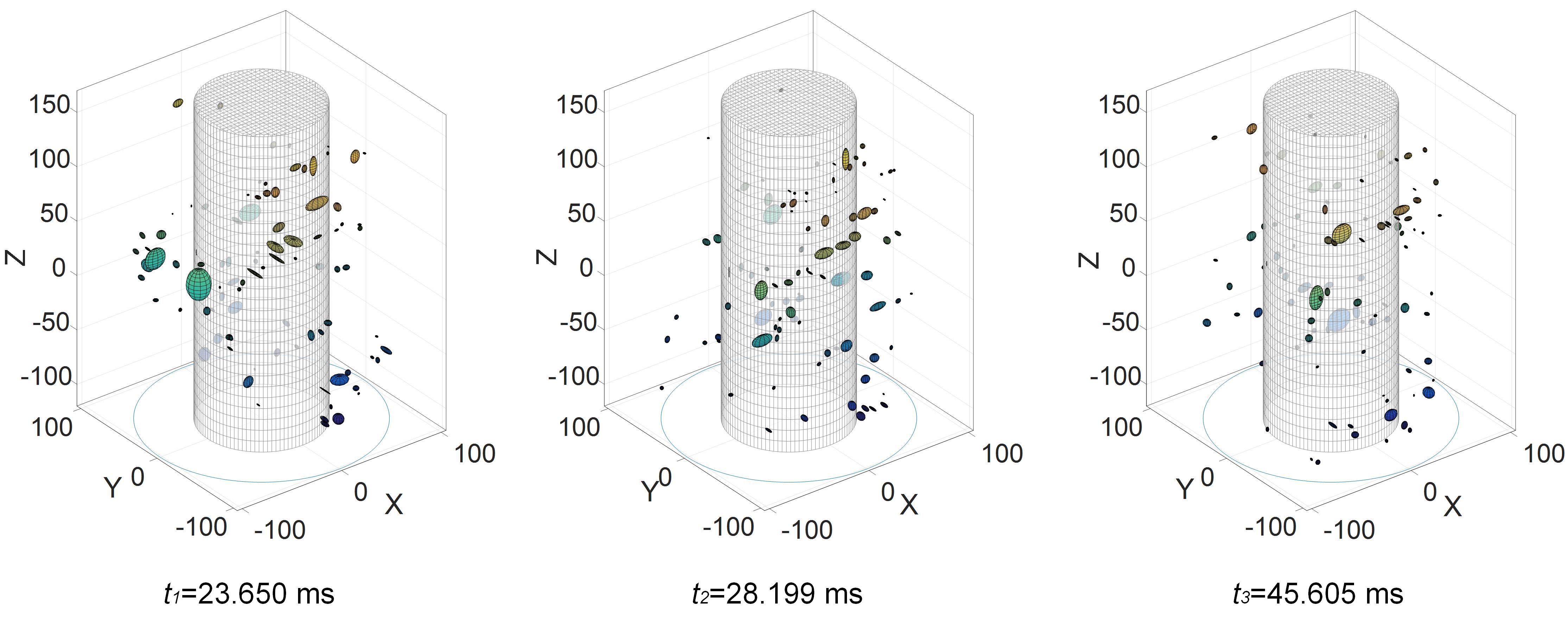}\\
  \caption{3-D reconstruction results of bubble flow in a short time}\label{fig06}
\end{figure}

An image processing algorithm called the frame differencing method was applied to detect the features of the cavitation bubbles. All the two dimensional (2D) bubble features of the relative motion compared to the previous near frame were extracted. The 3D bubble reconstruction in this paper was based on space rectangular coordinate system. The spatial coordinates of the cavitation bubbles in the valve was provided by the position information of the experiment images from the three directions of left (L), right (R), back (B). The bubble position coordinates on the horizontal axis from the L and B sides' images were directly used as the $x$ coordinate values of the bubbles in the spatial location and the R side's image provided the $y$ coordinate values. After matching the 2D bubble features, the 3D cavitation bubble cluster was reconstructed as shown in Figure \ref{fig06}. From the reconstruction results, the change process of the generation and collapse of the cavitation bubbles can be analyzed. And the number and motion rack of the bubbles can be further calculated.

\section{Conclusion}
In this paper, we proposed a novel 3D image capture method and applied an improved HS algorithm to optimize its design arrangement. At last, as a case study, this method was applied in a water hydraulic valve to capture and reconstruct the cavitation bubble cluster. By analyzing the position distribution and density of the Pareto solutions about a representative point, the advantage and improvement of the modified HS method was verified. Besides, the cavitation experiment was implemented. Through analysis and calculation of the experimental images, the 3D image of the cavitation bubbles was realized. The reconstruction results were quite accurate and effective. And the effectiveness of the 3D imaging device was testified. The 3D imaging method can be also applied in various multiphase flow measurements and has quite good flexibility and adaptability for different observed object.

\section*{Acknowledgment}
This work was supported by the Natural Science Foundation of China [grant number 51875113]; Natural Science Foundation of the Heilongjiang Province of China [grant number F2016003]; 'Jinshan Talent' Zhenjiang Manufacture 2025 Leading
 Talent Project; and 'Jiangyan Planning' Project in Yangzhong City.

\section*{References}

\bibliography{mybibfile}

\begin{thebibliography}{10}
\expandafter\ifx\csname url\endcsname\relax
  \def\url#1{\texttt{#1}}\fi
\expandafter\ifx\csname urlprefix\endcsname\relax\def\urlprefix{URL }\fi
\expandafter\ifx\csname href\endcsname\relax
  \def\href#1#2{#2} \def\path#1{#1}\fi

\bibitem{ZHENG20105264}
S.~qing Zheng, Y.~Yao, F.~fei Guo, R.~shan Bi, J.~ya~Li, Local bubble size
  distribution, gas-liquid interfacial areas and gas holdups in an up-flow
  ejector, Chemical Engineering Science 65~(18) (2010) 5264 -- 5271.
\newblock \href {http://dx.doi.org/10.1016/j.ces.2010.06.027}
  {\path{doi:10.1016/j.ces.2010.06.027}}.

\bibitem{HONKANEN201025}
M.~Honkanen, H.~Eloranta, P.~Saarenrinne, Digital imaging measurement of dense
  multiphase flows in industrial processes, Flow Measurement and
  Instrumentation 21~(1) (2010) 25 -- 32.
\newblock \href {http://dx.doi.org/10.1016/j.flowmeasinst.2009.11.001}
  {\path{doi:10.1016/j.flowmeasinst.2009.11.001}}.

\bibitem{WU20132928}
B.~Wu, J.~Kang, W.~Han, Design of dammann grating based on the parallel
  recombination simulated annealing algorithm, Optik - International Journal
  for Light and Electron Optics 124~(17) (2013) 2928 -- 2931.
\newblock \href {http://dx.doi.org/10.1016/j.ijleo.2012.08.078}
  {\path{doi:10.1016/j.ijleo.2012.08.078}}.

\bibitem{Xueting}
T.~Xue, L.~Qu, B.~Wu, Matching and 3-d reconstruction of multibubbles based on
  virtual stereo vision, IEEE Transactions on Instrumentation and Measurement
  63~(6) (2014) 1639--1647.

\bibitem{Xueting2}
T.~Xue, Y.~Chen, P.~Ge, Multibubbles segmentation and characteristic
  measurement in gas-liquid two-phase flow, Advances in Mechanical Engineering
  5 (2013) 143939.
\newblock \href {http://dx.doi.org/10.1155/2013/143939}
  {\path{doi:10.1155/2013/143939}}.

\bibitem{YU2016120}
L.~Yu, B.~Pan, Single-camera stereo-digital image correlation with a
  four-mirror adapter: optimized design and validation, Optics and Lasers in
  Engineering 87 (2016) 120 -- 128.
\newblock \href {http://dx.doi.org/10.1016/j.optlaseng.2016.03.014}
  {\path{doi:10.1016/j.optlaseng.2016.03.014}}.

\bibitem{PAN201625}
B.~Pan, L.~Yu, Y.~Yang, W.~Song, L.~Guo, Full-field transient 3d deformation
  measurement of 3d braided composite panels during ballistic impact using
  single-camera high-speed stereo-digital image correlation, Composite
  Structures 157 (2016) 25 -- 32.
\newblock \href {http://dx.doi.org/10.1016/j.compstruct.2016.08.017}
  {\path{doi:10.1016/j.compstruct.2016.08.017}}.

\bibitem{YU201717}
L.~Yu, B.~Pan, Full-frame, high-speed 3d shape and deformation measurements
  using stereo-digital image correlation and a single color high-speed camera,
  Optics and Lasers in Engineering 95 (2017) 17 -- 25.
\newblock \href {http://dx.doi.org/10.1016/j.optlaseng.2017.03.009}
  {\path{doi:10.1016/j.optlaseng.2017.03.009}}.

\bibitem{Pan44412}
B.~Pan, D.~Wu, L.~Yu, Optimization of a three-dimensional digital image
  correlation system for deformation measurements in extreme environments,
  Appl. Opt. 51~(19) (2012) 4409--4419.
\newblock \href {http://dx.doi.org/10.1364/AO.51.004409}
  {\path{doi:10.1364/AO.51.004409}}.

\bibitem{Pan2018}
B.~Pan, L.~Yu, Q.~Zhang, Review of single-camera stereo-digital image
  correlation techniques for full-field 3d shape and deformation measurement,
  Science China Technological Sciences 61~(1) (2018) 2--20.
\newblock \href {http://dx.doi.org/10.1007/s11431-017-9090-x}
  {\path{doi:10.1007/s11431-017-9090-x}}.

\bibitem{app8071144}
M.~Li, M.~Li, G.~Han, N.~Liu, Q.~Zhang, Y.~Wang, Optimization analysis of the
  energy management strategy of the new energy hybrid 100
  using a genetic algorithm, Applied Sciences 8~(7).

\bibitem{app8020175}
W.~Shi, L.~Wang, Z.~Lu, Q.~Zhang, Application of an artificial fish swarm
  algorithm in an optimum tuned mass damper design for a pedestrian bridge,
  Applied Sciences 8~(2).

\bibitem{Zong600201}
Z.~W. Geem, J.~H. Kim, G.~Loganathan, A new heuristic optimization algorithm:
  Harmony search, SIMULATION 76~(2) (2001) 60--68.
\newblock \href {http://dx.doi.org/10.1177/003754970107600201}
  {\path{doi:10.1177/003754970107600201}}.

\bibitem{WANG20102826}
C.-M. Wang, Y.-F. Huang, Self-adaptive harmony search algorithm for
  optimization, Expert Systems with Applications 37~(4) (2010) 2826 -- 2837.
\newblock \href {http://dx.doi.org/10.1016/j.eswa.2009.09.008}
  {\path{doi:10.1016/j.eswa.2009.09.008}}.

\bibitem{RN1552}
K.~S, V.~K. D.M, Optimal planning of active distribution networks with hybrid
  distributed energy resources using grid-based multi-objective harmony search
  algorithm, Applied Soft Computing 67 (2018) 387--398.
\newblock \href {http://dx.doi.org/10.1016/j.asoc.2018.03.009}
  {\path{doi:10.1016/j.asoc.2018.03.009}}.

\bibitem{RN1542}
K.~Z. Gao, P.~N. Suganthan, Q.~K. Pan, T.~J. Chua, T.~X. Cai, C.~S. Chong,
  Pareto-based grouping discrete harmony search algorithm for multi-objective
  flexible job shop scheduling, Information Sciences 289 (2014) 76--90.
\newblock \href {http://dx.doi.org/10.1016/j.ins.2014.07.039}
  {\path{doi:10.1016/j.ins.2014.07.039}}.

\bibitem{RN1539}
F.~De~Paola, N.~Fontana, M.~Giugni, G.~Marini, F.~Pugliese, An application of
  the harmony-search multi-objective (hsmo) optimization algorithm for the
  solution of pump scheduling problem, Procedia Engineering 162 (2016)
  494--502.
\newblock \href {http://dx.doi.org/10.1016/j.proeng.2016.11.093}
  {\path{doi:10.1016/j.proeng.2016.11.093}}.

\bibitem{RN1550}
J.~L. Ponz-Tienda, A.~Salcedo-Bernal, E.~Pellicer, J.~Benlloch-Marco, Improved
  adaptive harmony search algorithm for the resource leveling problem with
  minimal lags, Automation in Construction 77 (2017) 82--92.
\newblock \href {http://dx.doi.org/10.1016/j.autcon.2017.01.018}
  {\path{doi:10.1016/j.autcon.2017.01.018}}.

\bibitem{RN1541}
K.~Gao, Y.~Zhang, A.~Sadollah, A.~Lentzakis, R.~Su, Jaya, harmony search and
  water cycle algorithms for solving large-scale real-life urban traffic light
  scheduling problem, Swarm and Evolutionary Computation 37 (2017) 58--72.
\newblock \href {http://dx.doi.org/10.1016/j.swevo.2017.05.002}
  {\path{doi:10.1016/j.swevo.2017.05.002}}.

\bibitem{PAN2010830}
Q.-K. Pan, P.~Suganthan, M.~F. Tasgetiren, J.~Liang, A self-adaptive global
  best harmony search algorithm for continuous optimization problems, Applied
  Mathematics and Computation 216~(3) (2010) 830 -- 848.
\newblock \href {http://dx.doi.org/10.1016/j.amc.2010.01.088}
  {\path{doi:10.1016/j.amc.2010.01.088}}.

\bibitem{RN1538}
X.~Dai, X.~Yuan, Z.~Zhang, A self-adaptive multi-objective harmony search
  algorithm based on harmony memory variance, Applied Soft Computing 35 (2015)
  541--557.
\newblock \href {http://dx.doi.org/10.1016/j.asoc.2015.06.027}
  {\path{doi:10.1016/j.asoc.2015.06.027}}.

\bibitem{RN1536}
I.~Amaya, J.~Cruz, R.~Correa, Harmony search algorithm: a variant with
  self-regulated fretwidth, Applied Mathematics and Computation 266 (2015)
  1127--1152.
\newblock \href {http://dx.doi.org/10.1016/j.amc.2015.06.040}
  {\path{doi:10.1016/j.amc.2015.06.040}}.

\bibitem{RN1557}
X.~Yuan, J.~Zhao, Y.~Yang, Y.~Wang, Hybrid parallel chaos optimization
  algorithm with harmony search algorithm, Applied Soft Computing 17 (2014)
  12--22.
\newblock \href {http://dx.doi.org/10.1016/j.asoc.2013.12.016}
  {\path{doi:10.1016/j.asoc.2013.12.016}}.

\bibitem{RN1544}
M.~Gheisarnejad, An effective hybrid harmony search and cuckoo optimization
  algorithm based fuzzy pid controller for load frequency control, Applied Soft
  Computing 65 (2018) 121--138.
\newblock \href {http://dx.doi.org/10.1016/j.asoc.2018.01.007}
  {\path{doi:10.1016/j.asoc.2018.01.007}}.

\bibitem{gao2009uni}
X.-Z. Gao, X.~Wang, S.~J. Ovaska, Uni-modal and multi-modal optimization using
  modified harmony search methods, International Journal of Innovative
  Computing, Information and Control 5~(10) (2009) 2985--2996.

\bibitem{Miettinen2008}
K.~Miettinen, Introduction to Multiobjective Optimization: Noninteractive
  Approaches, Springer Berlin Heidelberg, Berlin, Heidelberg, 2008, pp. 1--26.
\newblock \href {http://dx.doi.org/10.1007/978-3-540-88908-3_1}
  {\path{doi:10.1007/978-3-540-88908-3_1}}.

\bibitem{CHEN20113336}
Y.~Chen, X.~Zou, W.~Xie, Convergence of multi-objective evolutionary algorithms
  to a uniformly distributed representation of the pareto front, Information
  Sciences 181~(16) (2011) 3336 -- 3355.
\newblock \href {http://dx.doi.org/10.1016/j.ins.2011.04.004}
  {\path{doi:10.1016/j.ins.2011.04.004}}.

\bibitem{COELLOCOELLO2002193}
C.~A.~C. Coello, E.~M. Montes, Constraint-handling in genetic algorithms
  through the use of dominance-based tournament selection, Advanced Engineering
  Informatics 16~(3) (2002) 193 -- 203.
\newblock \href {http://dx.doi.org/10.1016/S1474-0346(02)00011-3}
  {\path{doi:10.1016/S1474-0346(02)00011-3}}.

\bibitem{Mateo2012}
P.~M. Mateo, I.~Alberto, A mutation operator based on a pareto ranking for
  multi-objective evolutionary algorithms, Journal of Heuristics 18~(1) (2012)
  53--89.
\newblock \href {http://dx.doi.org/10.1007/s10732-011-9156-4}
  {\path{doi:10.1007/s10732-011-9156-4}}.

\end{thebibliography}

\end{document}